\RequirePackage{fix-cm}
\documentclass[twocolumn, epjc3, comma, sort&compress, natbib]{svjour3}
\bibpunct{[}{]}{,}{n}{}{,} 

\journalname{Eur. Phys. J. C}

\usepackage{latexsym}
\usepackage{amsmath}
\usepackage{amssymb}
\usepackage{amsfonts}
\usepackage{CJKutf8}

\usepackage[mathscr,scaled=1.15]{urwchancal}
\DeclareFontFamily{OT1}{pzc}{}
\DeclareFontShape{OT1}{pzc}{m}{it}%
{<-> s * [1.15] pzcmi7t}{}
\DeclareMathAlphabet{\mathpzc}{OT1}{pzc}{m}{it}

\usepackage{color}

\usepackage{supertabular}
\usepackage{placeins}
\usepackage{epsfig}
\usepackage{graphicx}

\definecolor{purple}{rgb}{0.5,0,0.5}
\definecolor{blue}{rgb}{0.0,0,0.9}
\definecolor{prdblue}{rgb}{0.133,0.118,0.498}
\usepackage[colorlinks=true, pdfstartview=FitV, linkcolor=prdblue, citecolor= prdblue, urlcolor=prdblue]{hyperref}

\begin{document}
\begin{CJK*}{UTF8}{gbsn}

\title{$\,$\\[-6ex]\hspace*{\fill}{\normalsize{\sf\emph{Preprint nos}.\
NJU-INP 090/24}}\\[1ex]
Pion Boer-Mulders function using a contact interaction}

\author{Dan-Dan Cheng（程丹丹)\thanksref{NJU,INP}%
    $\,^{\href{https://orcid.org/0009-0009-6466-4483}{\textcolor[rgb]{0.00,1.00,0.00}{\sf ID}}}$
\and
        Zhu-Fang~Cui (崔著钫)\thanksref{NJU,INP}%
       $^{\href{https://orcid.org/0000-0003-3890-0242}{\textcolor[rgb]{0.00,1.00,0.00}{\sf ID}}}$
\and
    \mbox{Minghui~Ding (丁明慧)\thanksref{NJU,INP,HZDR}%
    $\,^{\href{https://orcid.org/0000-0002-3690-1690}{\textcolor[rgb]{0.00,1.00,0.00}{\sf ID}}}$}
\and
    Craig D. Roberts\thanksref{NJU,INP}%
       $^{\href{https://orcid.org/0000-0002-2937-1361}{\textcolor[rgb]{0.00,1.00,0.00}{\sf ID}},}$
\and
\\Sebastian M.~Schmidt\thanksref{HZDR,TUD}%
    $\,^{\href{https://orcid.org/0000-0002-8947-1532}{\textcolor[rgb]{0.00,1.00,0.00}{\sf ID}},}$
}

\authorrunning{Dan-Dan Cheng \emph{et al}.} 

\institute{School of Physics, Nanjing University, Nanjing, Jiangsu 210093, China \label{NJU}
           \and
           Institute for Nonperturbative Physics, Nanjing University, Nanjing, Jiangsu 210093, China \label{INP}
           \and
           Helmholtz-Zentrum Dresden-Rossendorf, Bautzner Landstra{\ss}e 400, D-01328 Dresden, Germany \label{HZDR}
           \and
            Technische Universi\"at Dresden, 01062 Dresden, Germany \label{TUD}
\\[1ex]
Email:
\href{mailto:phycui@nju.edu.cn}{phycui@nju.edu.cn} (ZFC);
\href{mailto:mhding@nju.edu.cn}{mhding@nju.edu.cn} (MD);
\href{mailto:cdroberts@nju.edu.cn}{cdroberts@nju.edu.cn} (CDR)
            }

\date{2024 September 16}

\maketitle

\end{CJK*}

\begin{abstract}
A symmetry preserving treatment of a vector\,$\otimes$\,vector contact interaction (SCI) is used as the basis for calculations of the two pion transverse momentum dependent parton distribution functions (TMDs); namely, that for unpolarised valence degrees-of-freedom and the analogous Boer-Mulders (BM) function. \linebreak
Amongst other things, the analysis enables the following themes to be addressed:
the quark current mass dependence of pion TMDs;
the impact of the gauge link model on the positivity constraint that bounds the BM function relative to the unpolarised TMD;
the equivalence of direct diagrammatic and light-front wave function TMD calculations;
and the size of the BM shift.
Interpreted astutely, these SCI results enable one to draw insightful pictures of pion TMDs.
\end{abstract}

\section{Introduction}
The $J^{PC}=0^{-+}$ pion is Nature's most fundamental Nambu-Goldstone boson \cite{Nambu:1960tm, Goldstone:1961eq, Nambu:2009zza}.  Notwithstanding that, it is also readily understood as a bound state built from a valence light quark and a valence light antiquark \cite{Maris:1997hd}: considering the $\pi^+$, one is looking at $u \bar d$.
Owing largely to this dichotomy, pion properties present the clearest window onto the emergence of a proton-size mass, $m_p$, in the Standard Model, \emph{viz}.\ emergent hadron mass (EHM)  \cite{Roberts:2021nhw, Binosi:2022djx, Ding:2022ows, Ferreira:2023fva, Raya:2024ejx}.
Consequently, measurements of such properties are a high priority at existing and next generation accelerator facilities \cite{Chen:2020ijn, Arrington:2021biu, Quintans:2022utc}.

If one chooses to quantise quantum chromodynamics (QCD) using light-front coordinates, then hadron light-front wave functions (LFWFs), which possess a probability interpretation, are directly accessible \cite{Brodsky:1997de}.
Alternatively, one can calculate such a LFWF via careful and appropriate projection of the hadron's Poincar\'e-covariant Bethe-Salpeter wave function \cite{Chang:2013pq}.
In either case, the complete pion LFWF depends on two variables: $x$, the light-front fraction of the pion's total momentum carried by a valence quark; and the valence quark's light-front transverse momentum, $k_\perp$.  (The valence antiquark is associated with $1-x$, $-k_\perp$.)
In these terms, the valence quark parton distribution function (DF), ${\mathpzc u}(x;\zeta)$, has long been a focus of study.
Naturally, being defined via the modulus squared of the LFWF \cite{Diehl:2000xz}, ${\mathpzc u}(x;\zeta) > 0$ on $x\in (0,1)$.
Moreover, it is a true number density, \emph{i.e}., ${\mathpzc u}(x;\zeta) dx$ describes the number of valence quark partons carrying a light-front momentum fraction between $x$ and $x+dx$ at probe energy scale $\zeta$.
The pointwise behaviour of this function has been a source of controversy for thirty-five years \cite{Cui:2021mom, Lu:2023yna}.

Working within the context of, \emph{e.g}., generalised parton correlation functions \cite{Meissner:2008ay}, the  valence quark DF may be viewed as the integral of an associated transverse momentum dependent (TMD) DF:
\begin{equation}
 {\mathpzc u}(x;\zeta) = \int d^2 k_\perp f_1(x,k_\perp^2;\zeta) \,.
\end{equation}
This TMD is a $1+2$-dimensional number density, which expresses helicity-independent $x$-$k_\perp^2$ correlations in the pion valence quark LFWF as measured by a vector (photon) probe.
For instance, following Ref.\,\cite{Zhang:2020ecj}, then at the hadron scale, $\zeta=\zeta_{\cal H}$, where valence degrees-of-freedom can be argued to carry all pion properties \cite{Yin:2023dbw}, one may expect that $f_1(x,k_\perp^2;\zeta_{\cal H})$ peaks at $(x,k_\perp^2)=(1/2,0)$ and falls as either variable shifts away from this location; and at any fixed $x$, $f_1(x,k_\perp^2;\zeta)$ vanishes as $k_\perp^2/m_p^2 \to \infty$.
(Here, the rapidity scale dependence of the TMD is not shown explicitly.  In practical applications, it is often chosen to be the same as $\zeta$ \cite{Bacchetta:2017gcc}.)

\begin{figure}[t]
\centerline{%
\includegraphics[clip, width=0.40\textwidth]{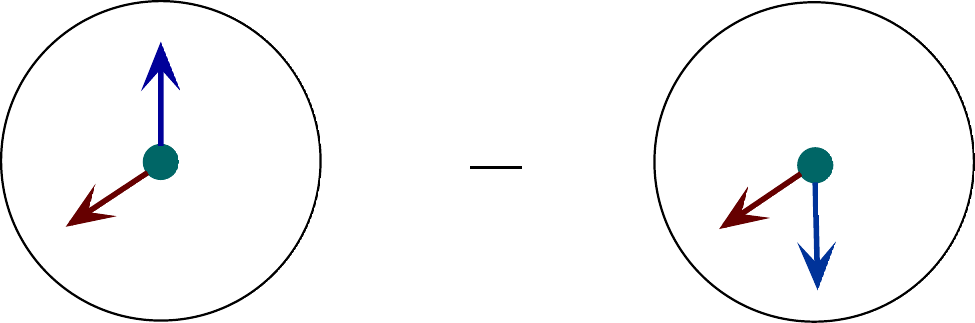}}
\caption{\label{FBM}
Number density interpretation of the Boer-Mulders function.
Legend.  Vertical blue arrows -- transverse polarisation of the quark, $\vec{S}$;
oblique red vectors -- quark transverse momentum vector, $\vec{k}_\perp$. }
\end{figure}

\begin{figure*}[t]
\centerline{%
\includegraphics[clip, width=0.775\textwidth]{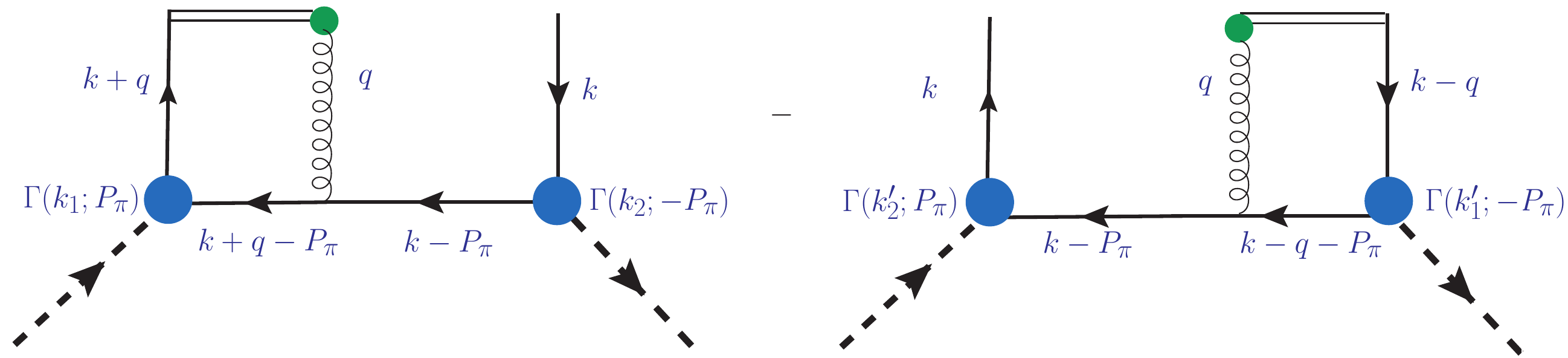}}
\caption{\label{ImageBM}
A nonzero Boer-Mulders function requires, at least, that the (valence) degree-of-freedom involved in the forward scattering event subsequently/initially interacts with the spectator via (multiple) gluon exchanges.
Legend.  Double line -- leading eikonal approximation to quark propagation under the influence of the gauge link, $1/n\cdot q$, where $n$ is a lightlike vector;
small green circle -- eikonal quark-gluon vertex, $[-g n_\mu]$, where $g$ is the strong coupling;
large blue  circle -- pion Bethe-Salpeter amplitude, $\Gamma_\pi(k_{\pm q,0};P)$, with $k_{\pm q}=k\pm q-P_\pi/2$, $k_0=k-P_\pi/2$;
thin solid line -- valence quark propagators, $S(k)$;
and spring-like line -- gluon propagator, $D_{\mu\nu}(q)$.
The relative negative sign between the two diagrams expresses the sign-change between initial- and final-state eikonal-quark interactions.
%
}
\end{figure*}

Given that the pion's valence degrees-of-freedom are $J=1/2$ fermions, another correlation is possible.  Namely, measured with respect to the pion $3$-momentum, it may be that the number density distribution of valence quark transverse spins depends on the quark's transverse momentum and that this dependence can be revealed by a vector probe -- see Fig.\,\ref{FBM}.  The Boer-Mulders (BM) function, $h_1(x,k_\perp^2)$, describes this correlation \cite{Boer:1997nt}.  It can be nonzero so long as, in calculating the associated $\gamma \pi \to \gamma \pi$ matrix element, interactions between the spectator and the struck, subsequently hard, valence degree-of-freedom are explicitly included -- see Fig.\,\ref{ImageBM}.  
The connection to spin correlations within a hadron make the BM function a prominent focus of contemporary phenomenology and theory; and recognition of the pion's pivotal role in elucidating EHM and its consequences means that developing an understanding of the pion BM function has become especially important.  Consequently, numerous studies of pion transverse structure have been completed within the past decade or so -- see, e.g., Refs.\,\cite{Lu:2012hh, Pasquini:2014ppa, Wang:2017onm, Ahmady:2019yvo, dePaula:2020qna, Zhu:2023lst, Kou:2023ady}.  Herein, we contribute to this discussion.

The presentation is arranged as follows.
Section~\ref{Secqqcf} introduces the two pion TMDs, linking them to distinct projections of a particular quark + quark correlation function that may be expressed diagrammatically -- see Fig.\,\ref{ImageBM}.
Calculation of and results for these TMDs are described in Sec.\,\ref{CalcTMD}.
The foundation for the analysis therein is a symmetry preserving treatment of a vector\,$\otimes$\,vector contact interaction (SCI) \cite{GutierrezGuerrero:2010md, Roberts:2010rn, Roberts:2011wy, Wilson:2011aa}, which is detailed in an appendix.
Two definitions of the gauge link required in order to obtain a nonzero BM function are considered and special attention is paid to constraints on such completions imposed by the positivity bound \cite{Bacchetta:1999kz}.
An alternate, LFWF approach to TMD calculation is sketched in Sec.\,\ref{LFWFcalc}.
It is useful in making contact with some other studies and in further elucidating the character of the positivity bound.
Aspects of TMD evolution are discussed in Sec.\,\ref{TMDEvolve}, with a focus on the so-called Boer-Mulders shift; namely, the mean transverse $y$-direction momentum of $x$-direction polarised quarks
in the unpolarisable pion.
The study is summarised in Sec.\,\ref{epilogue}, which also presents a perspective.

\section{Pion Boer-Mulders Function}
\label{Secqqcf}
In developing calculations of pion internal transverse structure, it is common to begin with the following Dirac matrix valued quark + quark correlation function:
\begin{align}
\label{D-17}
\Phi_{ij}(x,\mathbf{k}_{T}) & =
\int \frac{d^4\xi}{(2\pi)^3} \delta(n\cdot \xi)
e^{ik\cdot\xi}
\langle \pi(P)|\bar{q}_j(0) \nonumber  \\
& \quad \times {\cal L}^{\bar n}(0,\infty) {\cal L}^{\bar n}(\infty,\xi)
q_i(\xi)| \pi(P)\rangle\,,
\end{align}
where $i,j$ are spinor (Dirac matrix) indices.
Here we express ourselves with a Minkowski metric, for ease of comparison with existing literature, \emph{e.g}., Ref.\,\cite{Lu:2016pdp}.
Thus:
$n$ is a lightlike four-vector, $n^2=0$;
$\bar n$ is its conjugate, $\bar n^2=0$, $n\cdot \bar  n = 1$;
$x=n\cdot k/n\cdot P$, with $k$ the momentum of the active quark and $P$ is the total momentum of the pion;
and $(O_{\mu\nu}^\perp k^\nu) = (0,\vec{k}_\perp,0)$,
$O_{\mu\nu}^\perp = g_{\mu\nu} - n_\mu \bar n_\nu - \bar n_\mu n_\nu$.
As usual, the correlator involves gauge links, $ {\cal L}^{\bar n}$, that ensure gauge invariance.
In Eq.\,\eqref{D-17}, the paths are \cite{Lu:2016pdp} a product of one term running along the negative light-cone between zero and infinity, ${\cal L}^{\bar  n}$, and another connecting these points along a transverse path, ${\cal L}^{\bar n {\rm T}}$.
If one uses the light-cone gauge, in which the $n\cdot A$ component of each gauge field is set to zero, then ${\cal L}^{\bar  n} \equiv 0$.
Alternatively, ${\cal L}^{\bar n {\rm T}}\equiv 0$ in Feynman gauge.

Shifting now to a Euclidean formulation -- see, \emph{e.g}., Ref.\,\cite[Sec.\,1]{Ding:2022ows} for an explanation, in which case, $n\cdot \bar n = -1$, $O_{\mu\nu}^\perp = \delta_{\mu\nu} +n_\mu \bar n_\nu + \bar n_\mu n_\nu$, etc.; considering applicable symmetries and transformation properties; and working at leading order in $1/Q^2$ (leading twist); one may in general rewrite Eq.\,\eqref{D-17} as follows:
\begin{align}
\Phi_{q/\pi}(x,\mathbf{k}_{\perp})  & =
\tfrac{1}{2} \left[ f_{1\pi}^q(x,k_\perp^2) i \gamma\cdot n \right.  \nonumber \\
& \left. \qquad + h_{1\pi}^{q\perp}(x,k_\perp^2)  \tfrac{1}{M}  \sigma_{\mu\nu} k_{\perp \mu} n_\nu
\right]\,, \label{expansion}
\end{align}
where $M$ is a dynamically generated mass scale that is characteristic of EHM.
When dealing with the pion, a sensible choice is the infrared value of the QCD dressed light quark mass \cite{Roberts:2021nhw}: $M \approx 4 f_\pi$, where $f_\pi$ is the pion leptonic decay constant -- see also, \emph{e.g}., \ref{Adq}.
As a corollary of EHM, $M$ is nonzero in the chiral limit: it is an order parameter for dynamical chiral symmetry breaking.

Connecting Eqs.\,\eqref{D-17}, \eqref{expansion}, it is plain that
\begin{subequations}
\label{TMDprojections}
\begin{align}
f_{1\pi}^q(x,k_\perp^2) & = {\rm tr} \tfrac{1}{2} i\gamma\cdot\bar n \Phi_{q/\pi}(x,\mathbf{k}_{T}) \,,
\label{TMDprojectionA}\\
\frac{k_\perp^2}{M^2} h_{1\pi}^{q\perp}(x,k_\perp^2) &
= {\rm tr} \tfrac{1}{2 M} \sigma_{\mu\nu} k_{\perp \mu} \bar n_\nu \Phi_{q/\pi}(x,\mathbf{k}_{T})
\label{TMDprojectionB}\,.
\end{align}
\end{subequations}
Many studies use $M\to m_\pi$, \emph{i.e}., the pion mass; however, we judge this to be ill-advised because $m_\pi \to 0$ in the chiral limit whereas $f_\pi$ does not.

The helicity-independent number density, $f_{1\pi}^q(x,k_\perp^2)$, is always nonzero; but if the gauge links are omitted from Eq.\,\eqref{D-17}, then the BM function vanishes, \linebreak $h_{1\pi}^{q\perp}(x,k_\perp^2)\equiv 0$.  This is the signal that interactions between the spectator of the initial scattering event and the degree-of-freedom struck by the probe are crucial to obtaining $h_{1\pi}^{q\perp}(x,k_\perp^2) \neq 0$.  Herein, following, \emph{e.g}., Ref.\,\cite{Lu:2012hh}, we realise such interactions via the processes sketched in Fig.\,\ref{ImageBM}.

At this point, referring to the material already introduced, the number density distribution of valence quarks whose polarisation is transverse to the hadron's momentum direction vector, $\hat P$, is conventionally defined as follows \cite{Bacchetta:2004jz}:
\begin{align}
f_{q^\uparrow/\pi}(x,\vec{k}_\perp) & = \tfrac{1}{2} f_{1\pi}^q(x,k_\perp^2) \nonumber \\
& \qquad - h_{1\pi}^{q\perp}(x,k_\perp^2)\tfrac{1}{2M} \hat P \times \vec{k}_\perp \cdot \vec{S}_q\,.
\end{align}
(For a pion, it is a good approximation to consider $P \parallel n$.)  Then the number density asymmetry in Fig.\,\ref{FBM} corresponds to the following difference:
\begin{align}
f_{q^\uparrow/\pi}(x,\vec{k}_\perp) &  - f_{q^\downarrow/\pi}(x,\vec{k}_\perp) \nonumber \\
& = - h_{1\pi}^{q\perp}(x,k_\perp^2) \tfrac{|\vec{k}_\perp|}{M} \sin( \phi_{k_\perp} - \phi_S)\,,
\end{align}
where the azimuthal angles are measured between the indicated transverse vector and $\hat{P}$.
Exploiting the requirements imposed by positivity of the defining matrix element, one readily arrives at the following pointwise positivity bound \cite{Bacchetta:1999kz}:
\begin{equation}
\label{EqPositive}
 |k_\perp h_{1\pi}^{q\perp}(x,k_\perp^2)/M| \leq f^q_{1\pi}(x,k_\perp^2)\,.
\end{equation}

\section{Diagrammatic Calculation}
\label{CalcTMD}
\subsection{General remarks}
One may calculate the pion BM function by giving mathematical expression to the diagrams in Fig.\,\ref{ImageBM}.  We approach this by specifying the quark + quark interaction, computing every element in the diagrams, then combining them to produce a numerical result.  For the interaction, we use the SCI formulated in Refs.\,\cite{GutierrezGuerrero:2010md, Roberts:2010rn, Roberts:2011wy, Wilson:2011aa}, which has since been used in a wide variety of studies -- see, for instance, Refs.\,\cite{Zhang:2020ecj, Raya:2021pyr, Gutierrez-Guerrero:2021rsx, Cheng:2022jxe, Xing:2022sor, Xing:2022mvk, Xing:2023pms}.
Whilst the SCI is not a precision tool, it does have many merits, \emph{e.g}.:
algebraic simplicity;
simultaneous applicability to a wide variety of systems and processes;
potential for revealing insights that connect and explain numerous phenomena;
and capacity to serve as a means of checking the validity of algorithms employed in calculations that depend upon high performance computing.
Moreover, today's applications are typically parameter-free.
A brief recapitulation of the SCI is provided in \ref{AppendixSCI}.

The SCI result for the helicity-independent pion TMD is reported elsewhere \cite[Sec.\,6.1]{Zhang:2020ecj}:
\begin{align}
f_{1\pi}(x,k_\perp^2)
& = \frac{N_c}{2\pi^3} {\cal N}_{EF} \left[ E_\pi
\frac{\bar{\mathpzc C}_2(\varsigma)}{\varsigma} \right. \nonumber \\
& \left.
\qquad + 3 \, {\cal N}_{EF}\,x(1-x) m_\pi^2
\frac{\bar{\mathpzc C}_3(\varsigma)}{\varsigma^2}\right] \,,
\label{EqpionTMDT2}
\end{align}
with
${\cal N}_{EF}=(E_\pi - 2 F_\pi)$
and
$\varsigma = k_\perp^2 + M^2 - x (1-x) m_\pi^2$.
The special functions are defined in Eq.\,\eqref{eq:Cn} and the constants $E_\pi$, $F_\pi$ express the SCI pion bound-state amplitude -- see Eq.\,\eqref{pionBSA}.

\begin{figure}[t]
\centerline{%
\includegraphics[clip, width=0.44\textwidth]{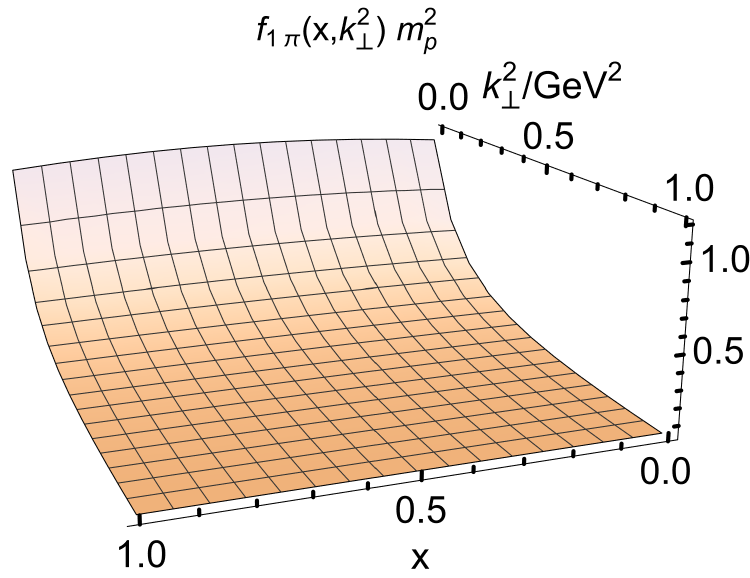}}
\caption{\label{Plotf1}
Hadron scale SCI result for the helicity-independent pion TMD drawn from Eq.\,\eqref{EqpionTMDT2}.}
\end{figure}

The helicity-independent SCI pion TMD is drawn in Fig.\,\ref{Plotf1}.
Recall that this is the result at the hadron scale, $\zeta_{\cal H}$, whereat valence degrees-of-freedom carry all pion properties \cite{Yin:2023dbw}.  In particular
\begin{equation}
\label{MomSum}
\int_0^1 dx \int d^2 {\vec k}_\perp 2 x f_{1\pi}(x,k_\perp^2;\zeta_{\cal H}) = 1\,,
\end{equation}
\emph{viz}.\ the light-front momentum sum rule is saturated by the pion's valence degrees-of-freedom.
(${\cal G}$-parity symmetry \cite{Lee:1956sw} is assumed herein.)
Evidently, the $\zeta_{\cal H}$ TMD is symmetric around $x=1/2$.
Further, the $k_\perp^2$ profile is almost independent of $x$ because $m_\pi^2/M^2 \ll 1$ -- \emph{cf}. Figs.\,\ref{Plotf1pis}, \ref{Ploth1pis} below.
One notes, too, that using the SCI, $f_{1\pi}(x,k_\perp^2)\neq 0$ is nonzero on $x\simeq 0,1$ at any finite $k_\perp^2$.
This is an artefact of the momentum-independent quark + quark interaction.
Using an interaction that becomes weaker with increasing momentum transfer, the hadron scale TMD vanishes at these endpoints \cite{Lu:2021sgg}.

Turning to the BM function, one can first use time-reversal invariance, expressed in the following operations,
\begin{equation}
\label{TRops}
\begin{array}{ll}
{\mathpzc T}^\dagger \gamma\cdot n^{\rm T} {\mathpzc T} = \gamma\cdot n \,, &
{\mathpzc T}^\dagger S(k)^{\rm T} {\mathpzc T} = S(k)\,, \\
{\mathpzc T}^\dagger \sigma_{ \alpha +}^{\rm T} {\mathpzc T} = -\sigma_{ \alpha +} \,, &
{\mathpzc T}^\dagger \Gamma_\pi(k;P)^{\rm T} {\mathpzc T} = \Gamma_\pi(k;-P)\,,
\end{array}
\end{equation}
where $\sigma_{\alpha +} = \sigma_{\alpha\beta} n_\beta$; and
${\mathpzc T}= \gamma_5 C$, with $C=\gamma_2\gamma_4$ being the charge-conjugation matrix,
to show that, in general, irrespective of the quark + quark interaction, the second diagram in Fig.\,\ref{ImageBM} maps into the first.  Consequently, the SCI yields:
\begin{align}
h_{1 \pi}^{\perp}(x, k_{\perp}^{2}) & \frac{ \vec{k}_{\perp \alpha}}{M}
 = N_{c} {\rm tr}_{\rm D}
\int \frac{d^{4}q}{(2 \pi)^{4}} \frac{d k_{3} dk_{4}}{(2 \pi)^{4}}
\delta_{n}^{x}(k)S(k-P _ {\pi}) \nonumber \\
& \times \Gamma (- P _ {\pi })S(k)\sigma_{ \alpha +} \frac { [-gn_{\mu }] }{n\cdot q} S(k +q)
\nonumber\\
&\times \Gamma( P_{\pi }) S(k+q- P_{\pi})[-g\gamma_{\nu}] D_{\mu \nu}(q) \,.
\label{FM-1}
\end{align}
Here the trace is over spinor indices;
$\delta_{n}^{x}(k)=\delta(n\cdot k-xn\cdot P _ {\pi})$;
$1/[n\cdot q]$ is the propagator of the eikonalised quark line and $[-gn_{\mu }]$ is the associated coupling to the gluon \cite{Collins:1989gx}, with $g$ being the strong coupling parameter;
and $D_{\mu \nu}$ is the gluon propagator that mediates the target + spectator interaction.

It is worth remarking that in a Euclidean metric formulation, the struck-quark on-shell condition for semi-inclusive deep inelastic scattering (SIDIS) is implemented via $1/[n\cdot q] \to - \pi \delta(n\cdot q)$, whereas $1/[n\cdot q] \to + \pi \delta(n\cdot q)$ for Drell-Yan (DY).  Using Eq.\,\eqref{FM-1}, it then follows that
\begin{equation}
h_{1 \pi}^{\perp}(x, k_{\perp}^{2})_{\rm SIDIS} =
- h_{1 \pi}^{\perp}(x, k_{\perp}^{2})_{\rm DY}.
\end{equation}

\subsection{SCI gauge link}
\label{SCIGL}
Working with the SCI, using Eqs.\,\eqref{KDinteraction}\,--\,\eqref{KCI} to write
\begin{equation}
g^2 D_{\mu\nu}(q) = \frac{4\pi \alpha_{\rm IR}}{m_G^2} \delta_{\mu\nu}
\end{equation}
then the SIDIS form of Eq.\,\eqref{FM-1} evaluates to the following expression:
\begin{align}
h_{1 \pi}^{\perp}(x,k_{\perp}^{2}) &=
- \frac{\alpha_{\rm IR}}{m_G^2} \frac{N_{c}}{4\pi^3}
{\cal N}_{EF} \frac{\bar{\mathpzc C}_2(\varsigma)}{\varsigma}\{[(E_\pi - F_\pi) M^2 \nonumber \\
&  \; - x(1-x) F_\pi m_\pi^2 ]\bar{\mathpzc C}_1(\varsigma_0)
+ F_\pi {\mathpzc C}_0(\varsigma_0)\}\,,
\label{hperpSCI}
\end{align}
with $\varsigma_0 = \left. \varsigma\right|_{k_\perp^2=0} = M^2 - x(1-x) m_\pi^2$.
Plainly:
the BM function is only nonzero because of the interaction between the eikonalised quark and the spectator -- this is highlighted by the $\alpha_{\rm IR}/m_G^2$ factor;
it is nonzero in the chiral limit, $m_\pi^2 =0$, so long as the pion is a Goldstone boson;
and the magnitude of the effect reflects the scale of EHM -- see the numerator factor of $M^2$.

\begin{figure}[t]
\centerline{%
\includegraphics[clip, width=0.44\textwidth]{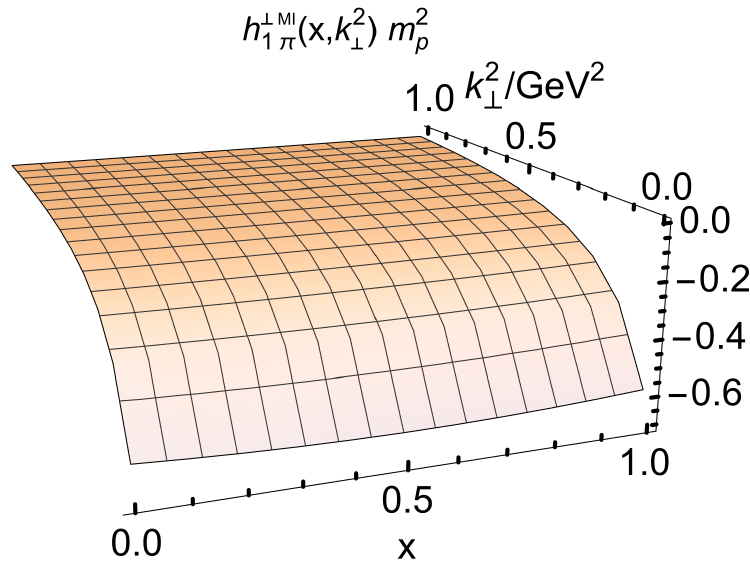}}
\caption{\label{Ploth1}
Hadron scale SCI result for the pion BM function, $h_{1 \pi}^{\perp}(x,k_{\perp}^{2};\zeta_{\cal H}) $, drawn from Eq.\,\eqref{hperpSCI}. Here, the gluon propagator used to calculate the gauge link is momentum-independent (MI).}
\end{figure}

The SCI BM function is depicted in Fig.\,\ref{Ploth1}.
Once again, we stress that this is the result at the hadron scale, $\zeta_{\cal H}$, whereat valence degrees-of-freedom carry all pion properties \cite{Yin:2023dbw}.
Compared with $f_{1\pi}(x,k_\perp^2)$ in Fig.\,\ref{Plotf1}, the BM function has a similar profile.  In fact, apart from an overall normalisation, the $(x,k_\perp^2)$-dependence is almost identical.
All other remarks made in the paragraph following Eq.\,\eqref{EqpionTMDT2} are also applicable here.

\subsection{Momentum-dependent link completion}
\label{MomDepLink}
Implemented as in Sec.\,\ref{SCIGL}, the gauge link contribution to $h_{1 \pi}^{\perp}$ is practically undamped.  For comparison, therefore, we also consider a more realistic example, \emph{viz}.\
\begin{equation}
\label{momentumglue}
g^2 D_{\mu\nu}(q) = \delta_{\mu\nu} \frac{4\pi \alpha_{\mathpzc L}}{q^2 + m_G^2}\,.
\end{equation}
If one uses $\alpha_{\mathpzc L} = \alpha_{\rm IR}$, then Eq.\,\eqref{momentumglue} reproduces the SCI form at infrared momenta.  On the other hand, for any value of $\alpha_{\mathpzc L}$,  Eq.\,\eqref{momentumglue} provides damping in the ultraviolet.  In this case:
\begin{align}
h_{1 \pi}^{\perp}&(x, k_{\perp}^{2}) =
- \alpha_{\mathpzc L} \frac{N_{c}}{4\pi^3}{\cal N}_{EF} \frac{\bar{\mathpzc C}_2(\varsigma)}{\varsigma} \nonumber \\
&
\times \int_{0}^{1} d\upsilon \big\{  2(1-\upsilon)
( E_{\pi} M^2 - F_{\pi} [M^2
 \nonumber \\
& \quad  + x(1-x)m_{\pi}^2 + \upsilon k_{\perp}^{2} ])
\frac{\bar {\mathpzc C}_2(\tilde\varsigma)}{\tilde\varsigma}
+ F_\pi \bar{\mathpzc C}_1(\tilde\varsigma) \big\}\,,
\label{PionBM}
\end{align}
where
$\tilde\varsigma =
\upsilon(1-\upsilon) k_{\perp}^{2} + (1 - \upsilon)\varsigma_0+\upsilon m_{G}^2$.

\begin{figure}[t]
\centerline{%
\includegraphics[clip, width=0.44\textwidth]{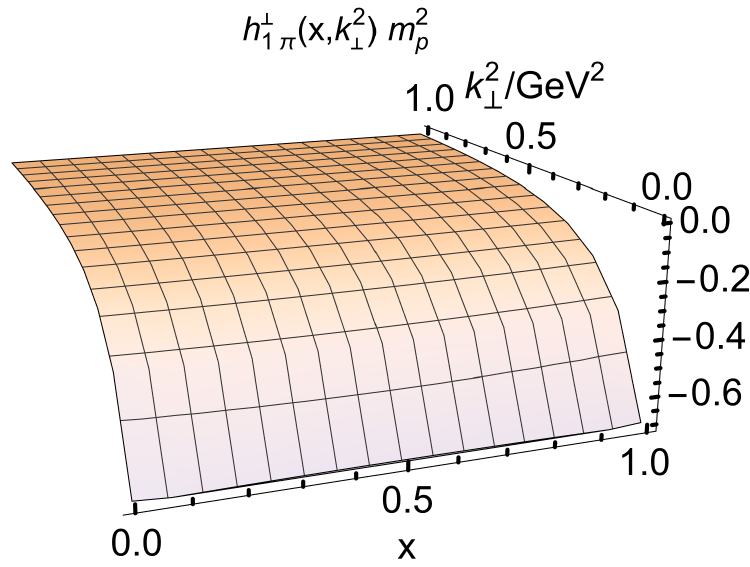}}
\caption{\label{Ploth1q2}
Hadron scale pion BM function, $h_{1 \pi}^{\perp}(x,k_{\perp}^{2}) $, drawn from Eq.\,\eqref{PionBM}. Here, the gluon propagator used to calculate the gauge link is momentum-dependent -- see Eqs.\,\eqref{momentumglue}, \eqref{alphaL}.
}
\end{figure}

It remains to choose a value for the coupling in Eq.\,\eqref{momentumglue}.
In the absence of additional robust constraints on the model for the gauge link contribution, we use
\begin{equation}
\alpha_{\mathpzc L} = 0.97 \pi \,,
\label{alphaL}
\end{equation}
\emph{viz}.\ the infrared value of QCD's process-independent effective charge \cite{Cui:2019dwv, Brodsky:2024zev}.  This value is known with a precision of 4\%.
Qualitatively, the BM function obtained from Eqs.\,\eqref{PionBM}, \eqref{alphaL} -- drawn in Fig.\,\ref{Ploth1q2} -- possesses all the properties listed after Eq.\,\eqref{hperpSCI}.

%

Other models for a momentum-dependent gauge link completion have been considered.
In contrast to Eq.\,\eqref{momentumglue}, they are often called ``nonperturbative'', being expressed in terms of a so-called lensing function -- see, \emph{e.g}., the discussion in Refs.\,\cite[Sec.\,IV]{Ahmady:2019yvo}, \cite[Sec.\,III\,B]{Kou:2023ady}.
However, such models do not necessarily represent an improvement because they typically lead to a mismatch between the bound-state kernel and the gauge link completion, which produces a violation of the positivity constraint in Eq.\,\eqref{EqPositive}.  This is discussed further in Sec.\,\ref{Quant} and following Eq.\,\eqref{PositivityLFWF}.  We do not employ such models herein.

\begin{figure}[t]
\vspace*{0.5ex}

\leftline{\hspace*{0.5em}{\large{\textsf{A}}}}
\vspace*{-1.5ex}
\includegraphics[width=0.44\textwidth]{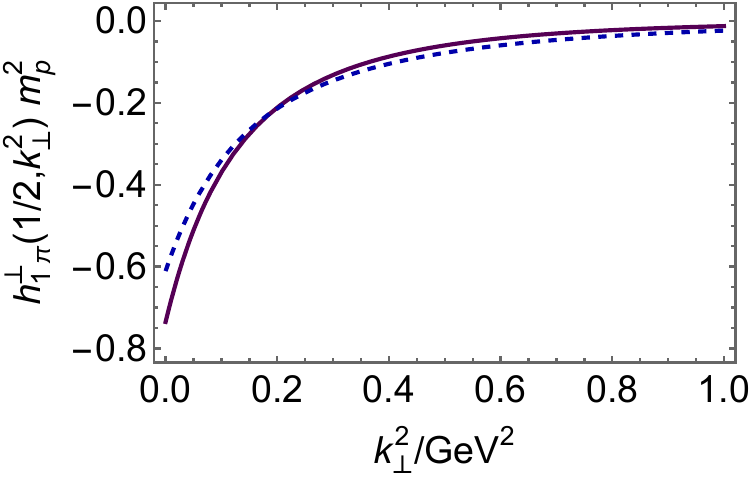}
\vspace*{+1ex}

\leftline{\hspace*{0.5em}{\large{\textsf{B}}}}
\vspace*{-1ex}
\includegraphics[width=0.46\textwidth]{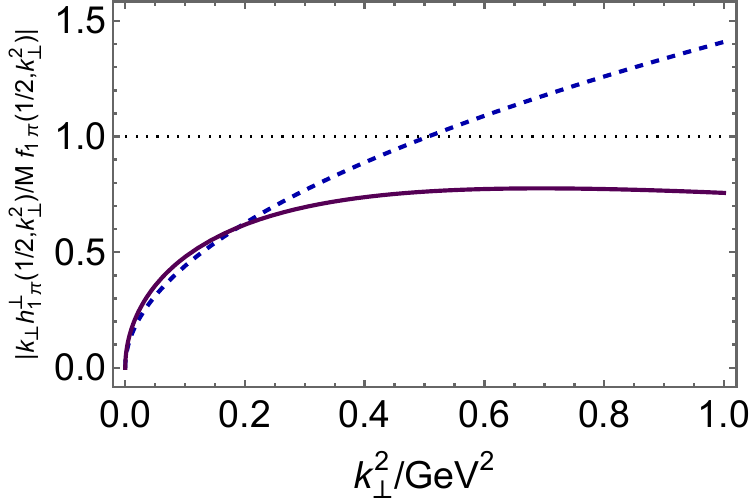}

\caption{\label{Ploth1cf}
{\sf Panel A}.
Comparison between the $k_\perp^2$ dependence of the curves in Figs.\,\ref{Ploth1}, \ref{Ploth1q2}, \emph{i.e}., $h_{1\pi}^{\perp {\rm MI}}(x=1/2,k_\perp^2)$ -- dashed blue \emph{cf}.\ $h_{1\pi}^{\perp}(x=1/2,k_\perp^2)$ solid purple.
{\sf Panel B}.
Positivity constraint, Eq.\,\eqref{EqPositive}: the bound is violated if the curve crosses the horizontal dotted line.
$|k_\perp h_{1\pi}^{\perp {\rm MI}}(x=1/2,k_\perp^2)/M f_{1\pi}(x=1/2,k_\perp^2)|$ -- dashed blue \emph{cf}.\ $|k_\perp h_{1\pi}^{\perp}(x=1/2,k_\perp^2)/M f_{1\pi}(x=1/2,k_\perp^2)|$ solid purple.}
\end{figure}

\subsection{Quantitative comparison}
\label{Quant}
Figure\,\ref{Ploth1cf}A compares the $k_\perp^2$ profiles of the pure SCI result for the BM function and that obtained using the interaction in Eq.\,\eqref{momentumglue} to define the gauge link.   The comparison is made using $x=1/2$; but with the SCI employed to define all other elements in the calculation, the result is practically the same for any value of $x$.  This image highlights the quicker decay with $k_\perp^2$ produced by the momentum-dependent gauge link interaction.
Naturally, the interaction used to define the gauge link has no impact on $f_{1\pi}(x,k_\perp^2)$, whereas the large $k_\perp^2$ behaviour of $h_{1\pi}^\perp(x,k_\perp^2)$ is very sensitive to this choice.

These observations lead one to consider the comparison in Fig.\,\ref{Ploth1cf}B; namely, to check the positivity bound described in connection with Eq.\,\eqref{EqPositive}.
Evidently, with a momentum-independent interaction used to calculate the gauge link contribution, positivity is violated.
Such an outcome is typical of treatments that provide greater support to the gauge link contribution than they do to the usual quark loop -- compare, \emph{e.g}., Refs.\,\cite{Lu:2004hu, Lu:2005rq}; and also the analyses in Refs.\,\cite{Pasquini:2014ppa, Wang:2017onm, Ahmady:2019yvo, Kou:2023ady}, which violate positivity for similar reasons.

On the other hand, using Eq.\,\eqref{momentumglue}, positivity is satisfied.  In this case, the quark and gluon propagators used throughout possess an ultraviolet momentum dependence that matches QCD expectations, up to logarithmic scaling violations, and the support of the usual quark loop is not curtailed without at least commensurate and consistent suppression of the gauge link range.

\begin{figure}[t]
\vspace*{0.5ex}

\leftline{\hspace*{0.5em}{\large{\textsf{A}}}}
\vspace*{-3ex}
\includegraphics[width=0.44\textwidth]{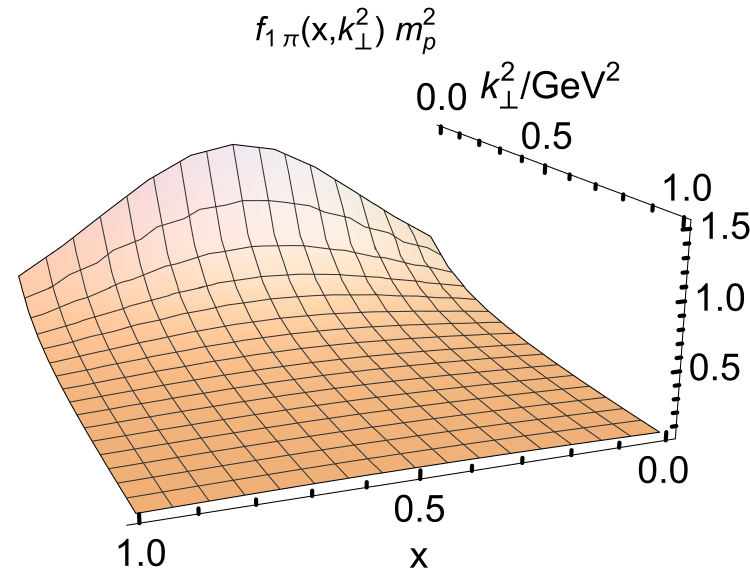}
\vspace*{+1ex}

\leftline{\hspace*{0.5em}{\large{\textsf{B}}}}
\vspace*{-1ex}
\includegraphics[width=0.46\textwidth]{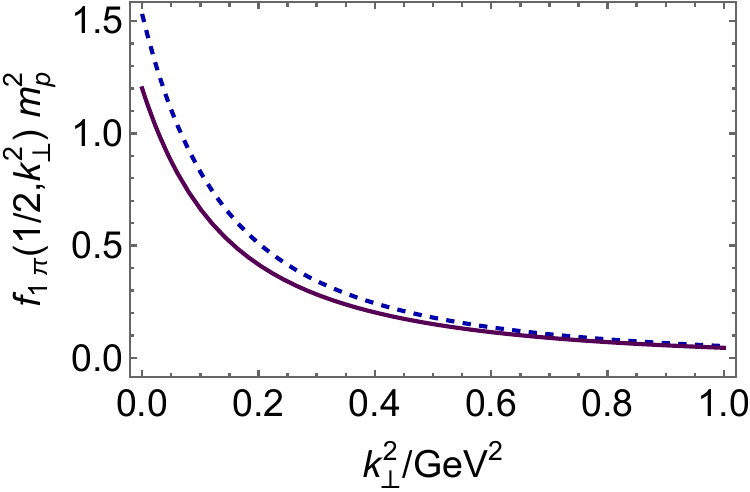}

\caption{\label{Plotf1pis}
SCI result for the helicity-independent $\pi_s$ TMD drawn from Eq.\,\eqref{EqpionTMDT2} using the inputs described in connection with Eq.\,\eqref{pionsBSA}.
{\sf Panel A}. TMD plotted as a function of $(x,k_\perp^2)$.
{\sf Panel B}. TMD at $x=1/2$ plotted as a function of $k_\perp^2$: solid purple curve -- $\pi$ result in Fig.\,\ref{Plotf1}; and dashed blue curve -- $\pi_s$ result.
}
\end{figure}

\subsection{Current mass dependence}
\label{Fictitious}
It is worth exposing the response of pion TMDs to an increase in quark current mass.  Therefore, consider a (fictitious) pseudoscalar state, $\pi_s$, composed of a quark and antiquark, each with the $s$ quark current mass listed in Table~\ref{Tab:DressedQuarks}.  Using the SCI, such a system has the mass and Bethe-Salpeter amplitude described in association with Eq.\,\eqref{pionsBSA}.

The helicity-independent $\pi_s$ TMD is drawn in \linebreak Fig.\,\ref{Plotf1pis}A.
Compared with the $\pi$ result in Fig.\,\ref{Plotf1}, the $x$ profile at fixed $k_\perp^2$ is far less dilated, with greater concentration around $x=1/2$.  This is the typical outcome of increasing the meson mass \cite{Noguera:2015iia, Lu:2021sgg}.
The comparison drawn in Fig.\,\ref{Plotf1pis}B highlights the increased magnitude of the $\pi_s$ helicity-independent TMD at its global maximum with respect to that of the $\pi$ and the more rapid decrease with increasing $k_\perp^2$ on the infrared domain.

\begin{figure}[t]
\vspace*{0.5ex}

\leftline{\hspace*{0.5em}{\large{\textsf{A}}}}
\vspace*{-3ex}
\includegraphics[width=0.44\textwidth]{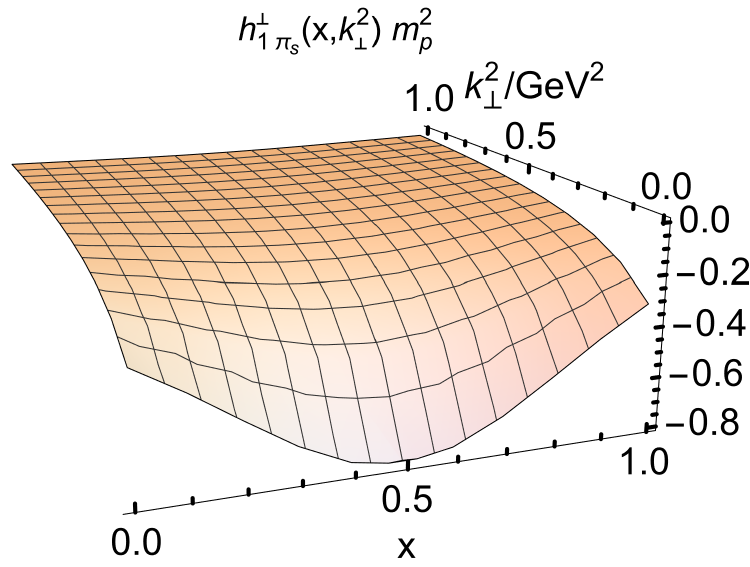}
\vspace*{+1ex}

\leftline{\hspace*{0.5em}{\large{\textsf{B}}}}
\vspace*{-1ex}
\includegraphics[width=0.46\textwidth]{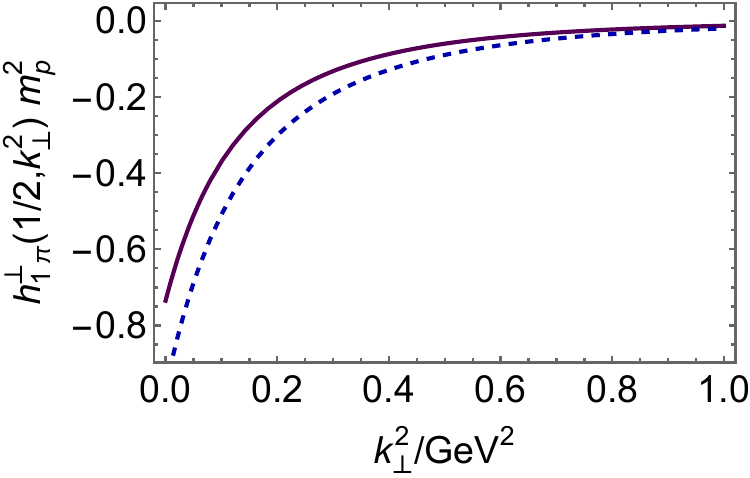}

\caption{\label{Ploth1pis}
SCI result for the $\pi_s$ BM function drawn from Eq.\,\eqref{PionBM} using the inputs described in connection with Eq.\,\eqref{pionsBSA} -- recall that Eq.\,\eqref{pionsBSA} was obtained using the momentum-dependent gluon propagator to define the gauge link contribution.
{\sf Panel A}. BM function as a function of $(x,k_\perp^2)$ \emph{cf}.\ Fig.\,\ref{Ploth1q2}.
{\sf Panel B}. BM function at $x=1/2$ plotted as a function of $k_\perp^2$: solid purple curve -- $\pi$ result in Fig.\,\ref{Ploth1cf}A; and dashed blue curve -- $\pi_s$ result.
}
\end{figure}

The complete helicity-dependent $\pi_s$ BM function is drawn in Fig.\,\ref{Ploth1pis}A, whereas Fig.\,\ref{Ploth1pis}B compares, at $x=1/2$, the $k_\perp^2$ profiles of the $\pi$ and $\pi_s$ BM functions.
Evidently, the response of the BM function to an increase in quark current mass is qualitatively equivalent to that of $f_{1\pi}(x,k_\perp^2)$.

\section{Calculation using Pion Light-Front Wave Function}
\label{LFWFcalc}
Suppose now that one has developed a light-front QCD Hamiltonian, using which a Fock space expansion may be defined whose leading two-particle element for the pion corresponds to the rainbow-ladder (RL) truncation bound state discussed, \emph{e.g}., in \ref{AppendixSCI}.
Namely, the LFWF associated with the leading term in the operator expansion of the Fock space is the pion's RL LFWF, denoted herein as $\Psi(x,k_\perp;\lambda_1,\lambda_2)$, where the second argument pair describes the helicities of the identified quasiparticles.
(An assumption of this sort is implicit in every light-front model of hadron structure.)

The creation and annihilation operators in this implicitly defined Fock space expansion are associated with quasiparticle Dirac spinors, for which Dirac-like equations are readily formulated and whose properties may therefrom be derived by explicit calculation.  Herein, it is only necessary to record that
\begin{subequations}
\begin{align}
\bar u(k,\lambda^\prime) i n\cdot \gamma & u(k,\lambda) = 2 n\cdot k \delta_{\lambda, \lambda^\prime} \,, \label{uTMDprojectionA}\\
\bar u(k^\prime,\lambda^\prime) & O^{\perp}_{\alpha\mu} k_\mu n_\nu \sigma_{\alpha\nu} u(k,\lambda)
\nonumber \\
& = -2 i \sqrt{n\cdot k^\prime n\cdot k} \lambda
\delta_{\lambda, -\lambda^\prime} |k_\perp| {\rm e}^{i \lambda \theta_\perp}\,,
\label{uTMDprojectionB}
\end{align}
\end{subequations}
where $\theta_\perp$ is the angle between $O^{\perp}_{\alpha\mu} k_\mu$ and $O^{\perp}_{\alpha\mu} k_\mu^\prime$ in the two-dimensional plane they define.
In this section, for ease of comparison with other studies and notwithstanding that we use a Euclidean metric, we implement normalisation conditions in line with those of Ref.\,\cite{Lepage:1980fj}.

{\allowdisplaybreaks
Projecting the SCI pion Bethe-Salpeter wave function onto the light-front, one obtains
\begin{subequations}
\label{piLFWF}
\begin{align}
\Psi(x,\vec{k}_\perp;\lambda_1,\lambda_2)  & = \psi(x,k_\perp^2) S_{\lambda_1,\lambda_2}(x,\vec{k}_\perp)\,,\\
\psi(x,k_\perp^2)  & = \frac{\sqrt{2 N_c}}{k_\perp^2+M^2-x(1-x)m_\pi^2}\,,
\label{PionSCILFWF}\\
S_{\lambda_1,\lambda_2}(x,\vec{k}_\perp) & =
\left[
\begin{array}{cc}
S_{\uparrow\uparrow} & S_{\uparrow\downarrow}\\
S_{\downarrow\uparrow} & S_{\downarrow\downarrow}
\end{array}
\right]\,,
\end{align}
with
\begin{align}
S_{\uparrow\uparrow} & = - {\cal N}_{EF} (k_1 - i k_2) \,,\\
S_{\uparrow\downarrow} & = \frac{E_\pi M^2 + F_\pi[k_\perp^2-M^2-x(1-x)m_\pi^2]}{M}\,,
\end{align}
\end{subequations}
$S_{\downarrow\downarrow} = S_{\uparrow\uparrow}^\ast $,
$S_{\downarrow\uparrow} = - S_{\uparrow\downarrow} $.
It is worth recalling here that the ultraviolet $k_\perp^2$ dependence of $\psi(x,k_\perp^2)$ in Eq.\,\eqref{PionSCILFWF} matches that of QCD, up to logarithmic corrections -- see Ref.\,\cite[Eq.\,(2.15)]{Lepage:1980fj}.
}

It is now possible to express our approximation to the Dirac matrix valued quark + quark correlation function in Eq.\,\eqref{D-17} using the pion RL LFWF:
\begin{align}
\label{D-17LFWF}
\Phi^{[{\mathpzc g}]}(x,\vec{k}_{\perp})  & =
\sum_{\lambda_1,\lambda_1^\prime,\lambda_2}
\int \frac{d^2 {k}_\perp^\prime}{16 \pi^3 \sqrt{n\cdot k n\cdot k^\prime}}
{\cal G}(x,\vec{k}_\perp,\vec{k}_\perp^\prime)
\nonumber \\
& \quad \times \psi(x,{k}_\perp^{\prime 2}) \psi(x,{k}_\perp^2)
[S_{\lambda_1^\prime,\lambda_2}(x,\vec{k}_\perp^\prime)]^\dagger
\nonumber \\
& \quad \times S_{\lambda_1,\lambda_2}(x,\vec{k}_\perp)\bar u(k^\prime,\lambda_1^\prime) \tfrac{1}{2}{\mathpzc g} u(k,\lambda_1)\,,
\end{align}
where ${\cal G}(x,\vec{k}_\perp,\vec{k}_\perp^\prime)$ is the model for the gauge link contribution and, anticipating Eqs.\,\eqref{TMDprojections}, we have included a contraction with some Dirac matrix ``${\mathpzc g}$''.

Considering the unpolarised TMD,
${\cal G}(x,\vec{k}_\perp,\vec{k}_\perp^\prime) \to \delta^2(\vec{k}_\perp-\vec{k}_\perp^\prime)$ in Eq.\,\eqref{D-17LFWF} because the gauge link plays no role.
Then, using Eqs.\,\eqref{TMDprojectionA}, \eqref{uTMDprojectionA}:
\begin{align}
& f_{1\pi}(x,k_\perp^2)  = \Phi^{[i \gamma\cdot \bar n]}(x,\vec{k}_{\perp}) \nonumber \\
& = \sum_{\lambda_1 \lambda_2 } \frac{1}{16\pi^3}
|\psi(x,{k}_\perp^2)|^2
[S_{\lambda_1,\lambda_2}(x,\vec{k}_\perp)]^\dagger
 S_{\lambda_1,\lambda_2}(x,\vec{k}_\perp) \nonumber \\
 & = \frac{N_c{\cal N}_{EF}}{4\pi^3}
 \left[
 \frac{E_\pi }{\varsigma}
 + \frac{{\cal N}_{EF} x(1-x)m_\pi^2}{\varsigma^2}
 +\frac{1}{{\cal N}_{EF}} \frac{F_\pi^2}{ M^2}
 \right]\,.
 \label{f1LFWF}
\end{align}

At this point, it is necessary to recall the symmetry preserving regularisation scheme used to define the contact interaction -- see \ref{AppendixSCI}.  Then one recognises that the final term within the parentheses in Eq.\,\eqref{f1LFWF} is associated with a diagrammatic quadratic divergence; hence, should be discarded.  Subsequently, using \linebreak Eq.\,\eqref{SCIdenominator}, Eq.\,\eqref{f1LFWF} is seen to reproduce Eq.\,\eqref{EqpionTMDT2}.

Thus prepared, we consider the BM function.
Writing ${\mathpzc g} = \sigma_{\mu\nu} k_{\perp \mu} \bar n_\nu/M$ in Eq.\,\eqref{D-17LFWF};
using \cite{Lu:2006kt}
\begin{equation}
{\cal G}(x,\vec{k}_\perp,\vec{k}_\perp^\prime)
= \frac{i \alpha}{2\pi}  D(q_\perp^2)\,,
\end{equation}
with $q_\perp = \vec{k}_\perp - \vec{k}_\perp^\prime$ and $D(q_\perp^2)$ a function to be specified;
and employing Eq.\,\eqref{uTMDprojectionB}, then
\begin{align}
& h_{1\pi}^{\perp}(x, k_\perp^2) = -\frac{N_c}{2\pi^3} \frac{{\cal N}_{EF}}{k_\perp^2}
\frac{\overline{\cal C}_2(\varsigma)}{\varsigma}
\int d^2 {k}_\perp^\prime \frac{\alpha_{\rm IR} D(q_\perp^2) }{2\pi[k_\perp^{\prime 2}+\varsigma_0]}  \nonumber \\
& \quad \times \left\{\rule{0em}{2.5ex}
k_\perp^2
( E_\pi M^2 + F_\pi [ k_\perp^{\prime 2}-M^2 -x(1-x) m_\pi^2 ] )
\right. \nonumber \\
& \;\quad \left.
-\vec{k}_\perp\cdot \vec{k}_\perp^\prime
( E_\pi M^2 + F_\pi [ k_\perp^2 -M^2 -x(1-x) m_\pi^2 ] )
\right\}\,.
\end{align}


The SCI result described above is recovered using $\alpha D(q_\perp^2) = \alpha_{\rm IR}/m_G^2$.
In this case, the integral involving $\vec{k}_\perp\cdot \vec{k}_\perp^\prime$ vanishes, owing to hyperspherical rotational invariance of the associated multiplicative factor.
The remaining integrals have the form described in \ref{AppendixSCI} -- see Eqs.\,\eqref{AppPower1}, \eqref{SCIdenominator}; and, evaluated, one recovers Eq.\,\eqref{hperpSCI}.

Naturally, the results in Sec.\,\ref{MomDepLink} are recovered with
\begin{equation}
\label{DqperpFree}
\alpha D(q_\perp^2) = \alpha_{\mathpzc L}/[m_G^2+q_\perp^2]\,.
\end{equation}

We have chosen to explain the LFWF path to TMDs because it straightforwardly enables contact to be made with some other models that produce a pion BM function.
For instance, one might consider the model obtained by setting $F_\pi \equiv 0$ in Eq.\,\eqref{piLFWF} -- a common oversimplification in pion models \cite{Chen:2012txa} -- and introducing
\begin{align}
\label{kp4LFWF}
\psi(x,k_\perp)  & = \frac{2 N_c E_\pi^2}{[k_\perp^2+M^2-x(1-x)m_\pi^2]^2}\,.
\end{align}
This LFWF yields
\begin{align}
f_{1\pi}(x,k_\perp^2)  & = \frac{3 N_c E_\pi^4}{2 \pi^3}
\left[\frac{1}{\varsigma^3} + \frac{x(1-x) m_\pi^2}{\varsigma^4}\right]
%
\end{align}
and, using Eq.\,\eqref{DqperpFree},
\begin{align}
h_{1\pi}^\perp(x,& k_\perp^2)  =
\frac{3 N_c E_\pi^4}{4\pi^3}\frac{M^2}{\varsigma}
\int_0^1 d\upsilon d\upsilon^\prime \upsilon \nonumber \\
& \times
\frac{1-\upsilon \upsilon^\prime}
{[1 - \upsilon \upsilon^\prime]\upsilon \upsilon^\prime k_\perp^2
+[1 - \upsilon \upsilon^\prime]\varsigma_0 + \upsilon \upsilon^\prime m_G^2
}\,. \label{PositivityLFWF}
\end{align}

In this case, positivity is violated.  Once again, that is because the pion LFWF in Eq.\,\eqref{kp4LFWF} cannot be produced by a one-gluon exchange interaction; hence, support in the quark loop is too compact in comparison with the range of the gauge link as prescribed by the gluon propagator in Eq.\,\eqref{DqperpFree}.  Phenomenologically, positivity can be restored by replacing Eq.\,\eqref{DqperpFree} by some suitably chosen, more rapidly damping form.  That, however, further weakens the connection between any such model and QCD.
Qualitatively, the models in Refs.\,\cite{Pasquini:2014ppa, Wang:2017onm, Ahmady:2019yvo, Kou:2023ady}
are members of the class illustrated by the above example.

\section{TMD Evolution and BM Shift}
\label{TMDEvolve}
A contemporary perspective on TMD evolution is presented elsewhere \cite{Boussarie:2023izj}.  Given a TMD, both the $x$ and $k_\perp^2$ profiles evolve, with the evolution equations containing elements which are essentially nonperturbative.  Today, that introduces practitioner-choice dependence into evolution outcomes.  This is especially true of the BM function.  Nevertheless, typically, evolution to larger (resolving and rapidity) scales leads to a shift of support in valence-quark TMDs toward $x=0$ and dilation of their $k_\perp^2$ profiles \cite{Wang:2017onm, Kou:2023ady}.

Instead of working with a complete TMD, ${\mathpzc f}(x,k_\perp^2)$, some simplicity is recovered by considering moments of the following form (here $k_\perp = |\vec{k}_\perp|$):
\begin{align}
{\mathpzc f}^{[j](n)}(b_\perp^2;\zeta) & =
\frac{n!}{m_\pi^{2n}}
\int_0^1 \! dx \, x^j \int d k_\perp k_\perp  \nonumber \\
& \times \quad  \left[\frac{k_\perp}{b_\perp }\right]^n
J_n(b_\perp k_\perp){\mathpzc  f}(x,k_\perp^2)\,,
\end{align}
where $J_n$ is a Bessel function of the first kind and $\zeta$ is the resolving scale.  In particular, using low-order such moments, the so-called generalised BM shift has been computed using lattice-regularised QCD (lQCD) \cite{Engelhardt:2015xja}:
\begin{equation}
\label{genBMshift}
\langle k_\perp \rangle_{\rm UT}(b_\perp^2;\zeta_2)
= m_\pi \frac{\tilde h_{1\pi}^{\perp[0](1)}(b_\perp^2;\zeta_2)}{f_{1\pi}^{[0](0)}(b_\perp^2;\zeta_2)}\,,
\end{equation}
where $\zeta_2 = 2\,{\rm GeV}$.  In the limit $b_\perp^2\to 0$, this ratio becomes the usual BM shift, \emph{viz}.\ the mean transverse $y$-direction momentum of $x$-direction polarised quarks in the unpolarisable pion.
In this section, $\tilde h_{1\pi}^{\perp} = (m_\pi/M) h_{1\pi}^{\perp}$, \emph{i.e}., we rescale our BM function by $m_\pi/M$ so as to match the convention used in the comparison studies.

In the limit $b_\perp^2\to 0$, the Eq.\,\eqref{genBMshift} denominator is unity, independent of the resolving scale:
\begin{equation}
f_{1\pi}^{[0](0)}(b_\perp^2=0;\zeta)=1\,.
\end{equation}

Regarding the numerator in Eq.\,\eqref{genBMshift}, consider the neighbourhood $b_\perp^2 \simeq 0$, on which $J_1(b_\perp k_\perp) = b_\perp k_\perp/2$, and define
\begin{align}
\tilde h_{1\pi}^{\perp(1)}(x;\zeta) & =
\frac{1}{2 m_\pi^{2}}
\int d k_\perp k_\perp k_\perp^2 \tilde h_{1\pi}^\perp(x,k_\perp^2)\,.
\end{align}
This object is directly related to $\tilde h_{1\pi}^{\perp[0](1)}$.  It is also proportional to a particular twist-three gluon + quark correlation function, $T_{q,F}^{(\sigma)}(x,x;\zeta)$, whose leading-order $\zeta$ evolution has been computed \cite{Kang:2012em}.
Various calculations have implemented the evolution equation therein with different truncations chosen for ease of use, \emph{viz}.\ the truncations can be expressed in terms of a simple splitting function.
Indeed, those we consider \cite{Pasquini:2014ppa, Wang:2017onm, Kou:2023ady} can be introduced by noting that they use leading-order evolution with splitting function:
\begin{subequations}
\label{splitting2}
\begin{align}
P_{qq}(z) & = P^{\Delta_T}_{qq}(z) + {\mathpzc s} \delta(1-z) \label{splitting1}\\
& = C_F \frac{2z}{(1-z)_+} + [C_F \tfrac{3}{2} + {\mathpzc s}] \delta(1-z) \,,
\end{align}
\end{subequations}
where $C_F = (N_c^2 - 1)/(2 N_c)$ and $\mathpzc s=0$ \cite{Pasquini:2014ppa} or $\mathpzc s= - N_c$ \cite{Wang:2017onm, Kou:2023ady}.  In Eq.\,\eqref{splitting1}, $P^{\Delta_T}_{qq}(z)$ is the splitting function for the chiral-odd proton transversity DF, whose correlator definition is similar to that of the pion BM function.  We proceed similarly herein, leaving consideration of the full evolution kernel to be discussed elsewhere.

From the perspective described above, the Mellin moments of $h_{1\pi}^{\perp(1)}(x;\zeta)$ are defined and evolve as follows \cite[Eq.\,(4.126)]{Ellis:1991qj}:
\begin{subequations}
\label{EvolveMellin}
\begin{align}
H_j(\zeta^2) & = \int_0^1 dx x^j \tilde h_{1\pi}^{\perp(1)}(x;\zeta) \,,\\
\zeta^2 \frac{d}{d\zeta^2} H_j(\zeta^2) & = \frac{\alpha(\zeta^2)}{2\pi} \gamma_h^j H_j(\zeta^2) \,,
\label{EvolutionHn}
\end{align}
\end{subequations}
where $\alpha(s)$ is a QCD running coupling.  Here, the anomalous dimensions are computed from the splitting function in Eq.\,\eqref{splitting2}:
\begin{equation}
\gamma_h^j  = \int_0^1 \! dz \, z^j P_{qq}(z) \,.
\end{equation}
The solution of Eq.\,\eqref{EvolveMellin} is:
\begin{equation}
    H_j(\zeta_2^2) = H_j(\zeta_1^2)
    \exp\left[ \gamma_h^m \int_{\zeta_1^2}^{\zeta_2^2} \!ds \,\alpha(s)/(2\pi s)]
    \right]\,;
    \label{EvolutionHn1}
\end{equation}
namely, given $H_j(\zeta_1^2)$, a moment at some scale $\zeta_1$, then its value at $\zeta_2$ is given by Eq.\,\eqref{EvolutionHn1}.

We interpret Eq.\,\eqref{EvolutionHn1} within the context of the all-orders (AO) evolution scheme described in Ref.\,\cite{Yin:2023dbw}, which has proved efficacious in numerous applications, \emph{e.g}., delivering unified predictions for all pion, kaon, and proton (unpolarised and polarised) DFs \cite{Cui:2020tdf, Chang:2022jri, Lu:2022cjx, Cheng:2023kmt, Yu:2024qsd}, and pion fragmentation functions \cite{Xing:2023pms}, that agree with much available data.
In this approach, $\alpha(s)$ is an effective charge \cite{Grunberg:1980ja, Grunberg:1982fw, Deur:2023dzc}, \emph{viz}.\ a QCD running coupling, defined such that when used to integrate the leading-order perturbative DGLAP equations \cite{Dokshitzer:1977sg, Gribov:1971zn, Lipatov:1974qm, Altarelli:1977zs}, it supplies an evolution scheme for all DFs -- both unpolarised and polarised, and for any hadron -- that is all-orders exact.  So defined, $\alpha(s)$ implicitly incorporates terms of arbitrarily high order in the perturbative coupling.  Any such effective charge has many valuable qualities, \emph{e.g}., it is: consistent with the QCD renormalisation group; renormalisation scheme independent; everywhere analytic and finite; and provides an infrared completion of any standard running coupling.

Within the AO scheme, Eq.\,\eqref{EvolutionHn1} can be rewritten as follows:
\begin{equation}
H_j(\zeta_2^2) = H_j(\zeta_1^2)
\left[\frac{\langle x \rangle_{{\mathpzc V}_{\pi}}^{\zeta_2}}{\langle x \rangle_{{\mathpzc V}_\pi}^{\zeta_1}},
\right]^{-\tfrac{3 \gamma_h^m}{4 C_F}},
\end{equation}
where $\langle x \rangle_{{\mathpzc V}_{\pi}}^\zeta$ is the sum of the light-front momentum fractions of the valence degrees-of-freedom in the pion at resolving scale $\zeta$.
Notably,  in the AO scheme, $\langle x \rangle_{{\mathpzc V}_\pi}^{\zeta_{\cal H}} = 1$ -- see Eq.\,\eqref{MomSum}.
This reexpression of Eq.\,\eqref{EvolutionHn1} highlights that evolution is largely insensitive to the pointwise form of $\alpha(k^2)$.  It is nevertheless worth remarking that the process-independent (PI) strong running coupling discussed in Refs.\,\cite{Cui:2019dwv, Brodsky:2024zev} has all the required properties.  A reliable interpolation of the PI charge is provided in Ref.\,\cite[Eq.\,(9)]{Ding:2022ows}.


The BM shift, $\langle k_\perp \rangle_{\rm UT}(b_\perp^2=0;\zeta_2)$,  is determined by the $j=0$ moment, for which one has
$\gamma_h^0 = {\mathpzc c} -C_F/2 $.
Using the one-loop perturbative running coupling for illustration, \emph{i.e}., writing
\begin{equation}
    \alpha(\zeta^2) \to \alpha_{\rm LO}(\zeta^2) = \frac{\pi \gamma_m}{\ln \zeta^2/\Lambda_{\rm QCD}^2}\,,
\end{equation}
with $\gamma_m = 12/(11 N_c - 2 n_f)$ and $n_f$ the number of active quark flavours, then one obtains
\begin{equation}
H_0(\zeta_2^2) = H_0(\zeta_1^2)
\left[\frac{\alpha_{\rm LO}(\zeta_2^2)}{ \alpha_{\rm LO}(\zeta_1^2)} \right]^{\tfrac{1}{4}(C_F-2{\mathpzc s})\gamma_m}.
\end{equation}
Clearly, for any ${\mathpzc s}<0$, the zeroth moment runs toward zero more quickly with increasing resolving scale than when ${\mathpzc s}=0$.  This observation reveals that the results in Refs.\,\cite{Wang:2017onm, Kou:2023ady} are inconsistent with the QCD evolution equation.

\begin{figure}[t]
\centerline{%
\includegraphics[clip, width=0.46\textwidth]{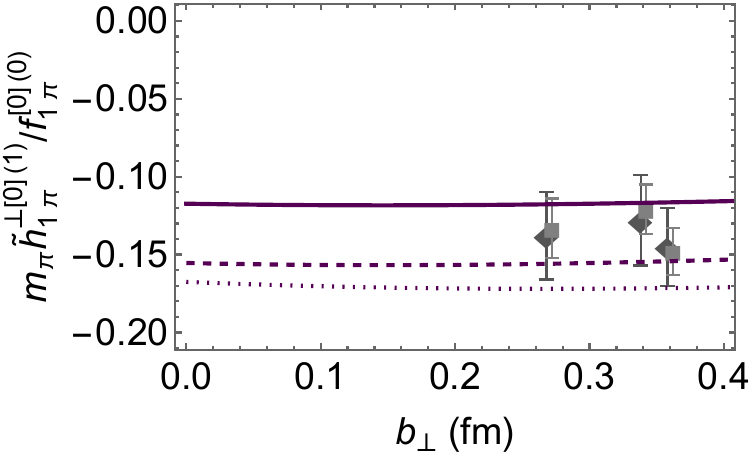}}
\caption{\label{PlotGshift}
Generalised BM shift, Eq.\,\eqref{genBMshift}.
Dotted purple curve -- hadron scale physical pion mass result;
dashed purple curve -- hadron scale heavy pion ($m_\pi=0.518\,$GeV);
and solid purple curve -- heavy pion result evolved $\zeta_{\cal H}\to \zeta_2$ using Eq.\,\eqref{EvolutionHn1}.
The points are available lQCD results \cite{Engelhardt:2015xja}, obtained with $m_\pi = 0.518\,$GeV at a resolving scale $\zeta=\zeta_2=2\,$GeV.
}
\end{figure}

Figure~\ref{PlotGshift} depicts the generalised BM shift as obtained using the momentum-dependent gauge link completion described in Sec.\,\ref{MomDepLink}.  As noted above, our direct calculations deliver TMDs at the hadron scale, $\zeta_{\cal H}$.  The dotted-purple curve is then the hadron-scale shift evaluated at the SCI physical pion mass.  It is practically $b_\perp^2$-independent.

Figure~\ref{PlotGshift} also displays lQCD results for the generalised BM shift, which were obtained with pion mass $m_\pi=0.518\,$GeV \cite{Engelhardt:2015xja}.
Using the SCI, such a pion mass is achieved with $M = 0.466\,$GeV, $E_\pi = 3.875$, $F_\pi=0.640$.
The dashed-purple curve in Fig.\,\ref{PlotGshift} is our hadron-scale prediction for the shift obtained with this ``heavy'' pion.  As also reported elsewhere \cite{Noguera:2015iia}, the magnitude of the shift decreases slowly with increasing meson mass.

The remaining curve in Fig.\,\ref{PlotGshift} is the heavy-pion result evolved $\zeta_{\cal H} \to \zeta_2$ using the $m=0$ version of Eq.\,\eqref{EvolutionHn1} and $\mathpzc s=0$:
\begin{equation}
H_0(\zeta_2^2)/H_0(\zeta_{\cal H}^2) = 0.76\,.
\end{equation}
Evidently, the impact of evolution is noticeable, but not dramatic; within existing uncertainties, the result thereby obtained is consistent with the lQCD points.  The BM shift is
\begin{equation}
  \langle k_\perp \rangle_{\rm UT}(b_\perp^2=0;\zeta_2) = -0.84 m_\pi \,.
\end{equation}

On the other hand, if one were to use $\mathpzc s= - N_c$, as is claimed to have been employed in Refs.\,\cite{Wang:2017onm, Kou:2023ady}, then the evolution factor is $0.22$, \emph{i.e}., a factor $3.4$ smaller and consistency with the lQCD points is lost.

\section{Summary and Perspective}
\label{epilogue}
Using a symmetry preserving treatment of a vector\,$\otimes$\,vec\-tor contact interaction (SCI) as the basis, we delivered predictions for the two nonzero pion transverse momentum dependent parton distribution functions (TMDs), \emph{viz}.\ that for unpolarised valence degrees-of-freedom and the Boer-Mulders (BM) function, which describes correlations between valence degree-of-freedom transverse spins and transverse momentum [Fig.\,\ref{FBM}].
The merit of using a SCI is that all analyses are largely algebraic, so the formulae and results exhibit a high level of transparency.  This enables clear insights to be drawn; not just of the SCI outcomes themselves, but also regarding results obtained using more sophisticated frameworks through relevant comparisons.  Moreover, interpreted carefully, SCI results can be physically relevant.

Whilst the unpolarised TMD is always nonzero, the BM function can only appear when, in calculating the associated $\gamma \pi \to \gamma \pi$ matrix element, interactions between the spectator and the struck and subsequently highly energetic valence degree-of-freedom are explicitly included [Fig.\,\ref{ImageBM}].  Formally, such interactions are described by gauge-field links, for which, in practice, phenomenological models are developed, typically involving an eikonal approximation to quark propagation under the influence of the gauge link.  We explicitly considered two models [Sec.\,\ref{CalcTMD}]: one is SCI based; and the other involves momentum-dependent gluon exchange between the spectator and eikonalised quark.

Our analysis reveals that the magnitude of the BM function increases with the size of the dressed-quark mass; hence, the strength of spin-momentum correlations is a signal and measure of emergent hadron mass (EHM).  Moreover, somewhat correlated with this, the peak magnitude of the BM function also increases with meson mass, as do the rates of decrease away from the peak with increasing $x$, $k_\perp^2$.  Importantly, the pointwise positivity constraint [Eq.\,\eqref{EqPositive}] can only be satisfied when due care is given to ensuring consistency between the support domains of the interaction which binds the pion and that used to characterise the gauge link.  In our view, future studies should pay more attention to this point.

In order to draw connections between various approaches to TMD computation, we also repeated our analyses using pion light-front wave functions (LFWFs) built to ensure consistency with the diagrammatic SCI calculations [Sec.\,\ref{LFWFcalc}].  Amongst other things, this was useful in highlighting the positivity issue.

Aspects of TMD evolution are sketched in Sec.\,\ref{TMDEvolve}, with a focus on the so-called BM shift.  Using a common choice for the splitting function, we present the all-orders evolution equation for this shift.  It can be expressed in terms of the sum of light-front momentum fractions of the pion's valence degrees-of-freedom.  The resulting SCI-based prediction for the shift is consistent with existing results obtained using lattice-regularised QCD.

A useful extension of this study would be to properly explore the impact of different truncations of the evolution kernel for the BM shift on the magnitude of this effect.  This is underway.  Going further, an SCI analysis of kaon BM functions would reveal insights into the role of quark current masses in driving distortions of TMD profiles.  Similarly, one may build upon Ref.\,\cite{Yu:2024qsd} and use the SCI to compute proton BM functions, a study that has the potential to expose novel impacts of diquark correlations on nucleon structure.  Such studies could employ either direct diagrammatic calculation or work with appropriate LFWFs.  Both avenues are being explored.

\begin{acknowledgements}
We are grateful for useful discussions with J.\,Rodr\'{\i}guez-Quintero, C.\,Shi and H.-Y.\,Xing.
Work supported by:
National Natural Science Foundation of China (grant no.\,12135007);
Natural Science Foundation of Jiangsu Province (grant no.\,BK20220122);
and
Helmholtz-Zentrum Dresden-Rossendorf, under the High Potential Programme.
\end{acknowledgements}

\appendix

\section{Contact Interaction}
\label{AppendixSCI}
Details concerning the SCI are available in many sources.  The following material is extracted from Refs.\,\cite[Sec.\,2]{Xu:2021iwv}, \cite[Appendix~A.2]{Yu:2024qsd}.  As therein, we work at leading-order in the systematic, symmetry preserving truncation scheme for the continuum bound state problem introduced in Refs.\,\cite{Munczek:1994zz, Bender:1996bb}, which is called rainbow-ladder (RL) truncation.
At this level, the foundation for any continuum meson bound-state problem is the quark + antiquark scattering kernel, which can be written:
\begin{align}
\label{KDinteraction}
\mathscr{K}_{\alpha_1\alpha_1',\alpha_2\alpha_2'}  & = \tilde{\mathpzc G}(k^2) T^k_{\mu\nu} [i\gamma_\mu]_{\alpha_1\alpha_1'} [i\gamma_\nu]_{\alpha_2\alpha_2'}\,,
\end{align}
where $k = p_1-p_1^\prime = p_2^\prime -p_2$, with $p_{1,2}$, $p_{1,2}^\prime$ being, respectively, the initial and final momenta of the scatterers, and $k^2T_{\mu\nu}^k = k^2\delta_{\mu\nu} - k_\mu k_\nu$.

The defining piece is $\tilde{\mathpzc G}$.  Since a gluon mass-scale emerges in QCD \cite{Gao:2017uox, Cui:2019dwv}, $\tilde{\mathpzc G}$ is nonzero and finite at infrared momenta:
\begin{align}
\label{SimpG}
\tilde{\mathpzc G}(k^2) & \stackrel{k^2 \simeq 0}{=} \frac{4\pi \alpha_{\rm IR}}{m_G^2}\,,
\end{align}
with \cite{Cui:2019dwv, Deur:2023dzc}: $m_G \approx 0.5\,$GeV, $\alpha_{\rm IR} \approx \pi$.
The value of $m_G$ is preserved herein.  Further, exploiting the fact that a SCI does not support relative momentum between bound-state constituents, one may simplify the tensor in Eqs.\,\eqref{KDinteraction}:
\begin{align}
\label{KCI}
\mathscr{K}_{\alpha_1\alpha_1',\alpha_2\alpha_2'}^{\rm CI}  & = \frac{4\pi \alpha_{\rm IR}}{m_G^2}
 [i\gamma_\mu]_{\alpha_1\alpha_1'} [i\gamma_\mu]_{\alpha_2\alpha_2'}\,.
 \end{align}

Confinement is effected by including an infrared mass scale, $\Lambda_{\rm ir}$, when solving all equations relevant to bound-state problems \cite{Ebert:1996vx}.  This artifice eliminates quark + antiquark production thresholds \cite{Krein:1990sf}.  The usual choice is $\Lambda_{\rm ir} = 0.24\,$GeV\,$=1/[0.82\,{\rm fm}]$ \cite{GutierrezGuerrero:2010md}, \emph{i.e}., a confinement length scale commensurate with proton radii \cite{Cui:2022fyr}.

SCI integrals also require ultraviolet regularisation.  This breaks the link between ultraviolet and infrared scales that is a distinguishing feature of QCD.  Consequently, the associated ultraviolet mass-scale, $\Lambda_{\rm uv}$, becomes a physical parameter, which may be interpreted as an upper bound on the domain whereupon amplitudes describing the associated systems are practically momentum-independent.

\subsection{Dressed quarks}
\label{Adq}
{\allowdisplaybreaks
The SCI gap equation is
\begin{align}
\label{GapEqn}
S^{-1}(p)  & = i\gamma\cdot p +m \nonumber \\
& \quad + \frac{16 \pi}{3} \frac{\alpha_{\rm IR}}{m_G^2}
\int \frac{d^4q}{(2\pi)^4} \gamma_\mu S(q) \gamma_\mu\,,
\end{align}
where $m$ is the light-quark current-mass.  Implementing a Poincar\'e-invariant regularisation, the solution is
\begin{equation}
\label{genS}
S^{-1}(p) = i \gamma\cdot p + M\,.
\end{equation}
The quark's dynamically generated dressed mass, $M$, is obtained by solving
\begin{equation}
M = m + M\frac{4\alpha_{\rm IR}}{3\pi m_G^2}\,\,{\cal C}_0^{\rm iu}(M^2)\,.
\label{gapactual}
\end{equation}
Here
\begin{align}
\nonumber
{\cal C}_0^{\rm iu}(\sigma) &=
\int_0^\infty\! ds \, s \int_{\tau_{\rm uv}^2=1/\Lambda_{\textrm{uv}}^{2}}^{\tau_{\rm ir}^2=1/\Lambda_{\textrm{ir}}^{2}} d\tau\,{\rm e}^{-\tau (s+\sigma)}\\
& =
\sigma \big[\Gamma(-1,\sigma \tau_{\rm uv}^2) - \Gamma(-1,\sigma \tau_{\rm ir}^2)\big],
\label{eq:C0}
\end{align}
where $\Gamma(\alpha,y)$ is the incomplete gamma-function.  The ``iu'' superscript indicates that the function depends on both the infrared and ultraviolet regularisation scales.
Functions of the following type arise in SCI bound-state equations:
\begin{align}
%
%
%
n !\, \overline{\cal C}^{\rm iu}_n(\sigma) & = \Gamma(n-1,\sigma \tau_{\textrm{uv}}^{2}) - \Gamma(n-1,\sigma \tau_{\textrm{ir}}^{2})\,,
\label{eq:Cn}
\end{align}
${\cal C}^{\rm iu}_n(\sigma)=\sigma \overline{\cal C}^{\rm iu}_n(\sigma)$, $n\in {\mathbb Z}^\geq$.}

\begin{table}[t]
\caption{\label{Tab:DressedQuarks}
Coupling, $\alpha_{\rm IR}$, ultraviolet cutoff, $\Lambda_{\rm uv}$, and current-quark masses, $m$, that enable a good description of flavour-nonsinglet pseudoscalar meson properties, along with the dressed-quark masses, $M$, pion mass, $m_{\pi}$, and leptonic decay constant, $f_{\pi}$, they produce \cite{Xu:2021iwv}; all obtained with $m_G=0.5\,$GeV, $\Lambda_{\rm ir} = 0.24\,$GeV.
Empirically, at a meaningful level of precision \cite{ParticleDataGroup:2024cfk}:
$m_\pi =0.14$, $f_\pi=0.092$.
%
Also included: SCI current- and dressed-quark masses appropriate to strange ($s$) quarks.
(We assume isospin symmetry and list dimensioned quantities in GeV.)}
\begin{center}
\begin{tabular*}
{\hsize}
{
c@{\extracolsep{0ptplus1fil}}|
c@{\extracolsep{0ptplus1fil}}
c@{\extracolsep{0ptplus1fil}}
c@{\extracolsep{0ptplus1fil}}
c@{\extracolsep{0ptplus1fil}}
c@{\extracolsep{0ptplus1fil}}
|c@{\extracolsep{0ptplus1fil}}
c@{\extracolsep{0ptplus1fil}}
c@{\extracolsep{0ptplus1fil}}
c@{\extracolsep{0ptplus1fil}}}\hline
$\alpha_{\rm IR}\ $ & $\Lambda_{\rm uv}$ & $m$ & $m_s$  &   $M$ & $M_s\ $ & $m_{\pi}$ & $f_{\pi}$ & $m_{\pi_s}$ \\\hline
$1.13\phantom{2}$ & $0.91\ $ & $0.0068\ $ & $0.16\ $& 0.37$\ $ & $0.53\ $ & 0.14 & 0.10 & 0.69\   \\\hline
\end{tabular*}
\end{center}
\end{table}

It is worth remarking here that, practically, our regularisation scheme implements the following replacement:
\begin{subequations}
\label{AppPower1}
\begin{align}
\frac{\Gamma(\alpha)}{\varsigma^\alpha} & =  \int_0^\infty d\tau \tau^{\alpha-1} {\rm e}^{- \tau \varsigma} \\
& \to
\int_{\tau_{\rm uv}^2}^{\tau_{\rm ir}^2} d\tau \tau^{\alpha-1} {\rm e}^{- \tau \varsigma} \\
& =  \int_{\tau_{\rm uv}^2}^{\infty} d\tau \tau^{\alpha-1} {\rm e}^{- \tau \varsigma}
- \int_{\tau_{\rm ir}^2}^{\infty} d\tau \tau^{\alpha-1} {\rm e}^{- \tau \varsigma} \\
& =
\frac{1}{\varsigma^\alpha}
[ \Gamma(\alpha,\varsigma\tau_{\rm uv}^2) - \Gamma(\alpha,\varsigma \tau_{\rm uv}^2) ]\,;
\end{align}
\end{subequations}
namely,
\begin{align}
\frac{1}{\varsigma^\alpha}
& \to  \frac{1}{\varsigma^\alpha}\frac{\Gamma(\alpha+2)}{\Gamma(\alpha)}\,
\overline{\cal C}^{\rm iu}_{\alpha+1}(\varsigma)\,.
\label{SCIdenominator}
\end{align}

The SCI parameters were fixed in Ref.\,\cite{Xu:2021iwv}; and those relevant herein are reported in Table~\ref{Tab:DressedQuarks}.

\subsection{Pion bound state}
The $\pi$-meson emerges as a quark +anti\-quark bound-state, whose structure is described by a Bethe-Salpeter amplitude that takes the following form when using the SCI:
\begin{align}
\Gamma_{\pi}(P) = \gamma_5 \left[ i E_{\pi}(P) + \frac{1}{M}\gamma\cdot P F_{\pi}(P)\right]\,,
\label{PSBSA}
\end{align}
where $P$ is the pion total momentum, $P^2 = -m_{\pi}^2$.
The amplitude and $m_{\pi}^2$ are obtained by solving the following Bethe-Salpeter equation $(t_+ = t+P)$:
\begin{align}
\Gamma_{\pi}(P)  & =  - \frac{16 \pi}{3} \frac{\alpha_{\rm IR}}{m_G^2}
\int \! \frac{d^4t}{(2\pi)^4} \gamma_\mu S(t_+) \Gamma_{\pi}(P)S(t) \gamma_\mu \,.
\label{LBSEI}
\end{align}
As stressed in Refs.\,\cite{GutierrezGuerrero:2010md, Chen:2012txa}, the axial-vector Ward-Green-Takahashi identity is violated of one omits the $\gamma\cdot P F_{\pi}(P)$ term.

Implementation of the SCI exploits a di\-men\-sional-regularisation-like identity:
\begin{equation}
0 = \int_0^1d\alpha \,
\big[ {\cal C}_0^{\rm iu}(\omega(\alpha,P^2))
%
+ \, {\cal C}^{\rm iu}_1(\omega(\alpha,P^2))\big], \label{avwtiP}
\end{equation}
where ($\hat \alpha = 1-\alpha$)
\begin{align}
\omega(\alpha,P^2) &= M^2+ \alpha \hat\alpha P^2\,.
\label{eq:omega}
\end{align}

Working from Eq.\,\eqref{avwtiP}, one arrives at the following Bethe-Salpeter equation:
\begin{equation}
\label{bsefinalE}
\left[
\begin{array}{c}
E_{\pi}(P)\\
F_{\pi}(P)
\end{array}
\right]
= \frac{4 \alpha_{\rm IR}}{3\pi m_G^2}
\left[
\begin{array}{cc}
{\cal K}_{EE}^{\pi} & {\cal K}_{EF}^{\pi} \\
{\cal K}_{FE}^{\pi} & {\cal K}_{FF}^{\pi}
\end{array}\right]
\left[\begin{array}{c}
E_{\pi}(P)\\
F_{\pi}(P)
\end{array}
\right],
\end{equation}
with
{\allowdisplaybreaks
\begin{subequations}
\label{fgKernel}
\begin{eqnarray}
\nonumber
{\cal K}_{EE}^{\pi} &=&
\int_0^1d\alpha \bigg\{
{\cal C}_0^{\rm iu}(\omega( \alpha, Q^2))  \\
&& \quad - 2 \alpha \hat\alpha P^2
\overline{\cal C}^{\rm iu}_1(\omega(\alpha, Q^2))\bigg\},\\
{\cal K}_{EF}^{\pi} &=& P^2 \int_0^1d\alpha\, \overline{\cal C}^{\rm iu}_1(\omega(\alpha, P^2)),\\
{\cal K}_{FE}^{\pi} &=& \frac{M^2}{2}\int_0^1d\alpha\, \overline{\cal C}^{\rm iu}_1(\omega(\alpha, P^2)),\\
{\cal K}_{FF}^{\pi} &=& - 2 {\cal K}_{FE}^{\pi}\,.
\end{eqnarray}
\end{subequations}}

The value of $P^2=-m_{\pi}^2$ for which Eq.\,\eqref{bsefinalE} is satisfied corresponds to the bound-state mass, with the calculated result listed in Table~\ref{Tab:DressedQuarks}, and the associated solution vector is the pion's Bethe-Salpeter amplitude.

In the calculation of observables, the canonically normalised amplitude must be used, \emph{i.e}., the amplitude obtained after rescaling such that
\begin{equation}
\label{normcan}
1=\left. \frac{d}{d P^2}\Pi_{\pi}(Z,P)\right|_{Z=P},
\end{equation}
where, with the trace over spinor indices:
\begin{align}
\Pi_{\pi}(Z,Q) & = 6 {\rm tr}_{\rm D} \!\! \int\! \frac{d^4t}{(2\pi)^4}   \Gamma_{\pi}(-Z)
 S(t_+) \, \Gamma_{\pi}(Z)\, S(t)\,.
 \label{normcan2}
\end{align}
The dimensionless result is
\begin{equation}
\label{pionBSA}
E_\pi = 3.59\,,
\quad F_\pi = 0.47\,.
\end{equation}

In terms of the canonically normalised Bethe-Salpe\-ter amplitude, the pseudoscalar meson's leptonic decay constant is
\begin{align}
f_{\pi} &= \frac{N_c}{2\pi^2 M}\,
\big[ E_{\pi} {\cal K}_{FE}^{\pi} + F_{\pi}{\cal K}_{FF}^{\pi} \big]_{Q^2=-m_{\pi}^2}\,. \label{ffg}
\end{align}
With our normalisation, the empirical value of the pion's leptonic decay constant is $f_\pi =0.092\,$GeV \cite{ParticleDataGroup:2024cfk}.

In Sec.\,\ref{Fictitious} we report the TMDs of a (fictitious) pseudoscalar state composed of a quark and antiquark, each with the $s$ quark current mass listed in Table~\ref{Tab:DressedQuarks}.  This bound state has mass $m_{\pi_s} = 0.69\,$GeV and a Bethe-Salpeter amplitude characterised by
\begin{equation}
\label{pionsBSA}
E_{\pi_s} = 4.04\,,
\quad F_{\pi_s} = 0.74\,.
\end{equation}


\end{document}